\documentclass[11pt,a4paper]{article}
\usepackage[centertags]{amsmath}
\usepackage{amsfonts}
\usepackage{amssymb}
\usepackage{amsthm}
\usepackage{graphicx}
\usepackage{ccaption}

\captionnamefont{\bfseries}
  \captiontitlefont{\small\sffamily}
  \captiondelim{: }
  \hangcaption

\author{John Hammersley\footnote{John.Hammersley@dunelm.org.uk} \\ \\ \textit{Department of Mathematical Sciences,} \\ \textit{Durham University,
South Road, Durham DH1 3LE UK}}
\title{\vspace{-1cm} \begin{flushright} \footnotesize{DCPT-07/35}
\end{flushright} \vspace{1cm} A critical dimension for the stability of perfect fluid spheres of radiation}
\date{}
\begin{document}

\maketitle

\begin{abstract}
An analysis of radiating perfect fluid models with asymptotically
AdS boundary conditions is presented. Such scenarios consist of a
spherical gas of radiation (a ``star'') localised near the centre of
the spacetime due to the confining nature of the AdS potential. We
consider the variation of the total mass of the star as a function
of the central density, and observe that for large enough
dimensionality, the mass increases monotonically with the density.
However in the lower dimensional cases, oscillations appear,
indicating that the perfect fluid model of the star is becoming
unrealistic. We find the critical dimension separating these two
regimes to be eleven.
\end{abstract}

\section{Introduction}

There are numerous effects in physics which are dependent on the
dimension of the spacetime in which they live, and concepts which
have mathematically simple models when restricted to one or two
dimensions may become intractably complex as the dimensionality
increases. Furthermore, in some situations there is a sharp contrast
in the behaviour of the system when the number of dimensions passes
above or below a certain ``critical dimension''. In the field of
general relativity there are already some instances of such
dimension dependent phenomena; for example, the Gregory-Laflamme
instability \cite{greglaf1,greglaf2} of black strings, and the work
of Belinsky, Khalatnikov and Lifshitz (BKL) \cite{bkl1} and its
extensions \cite{dema1,dema2,elsk}, where the dynamics of a
spacetime in the vicinity of a cosmological singularity were
studied. In the latter case they found that the general behaviour of
the relevant Einstein solutions changed from ``chaotic'' in the low
dimensional cases to non-chaotic in higher dimensions. It is this
extra complexity and appearance of a critical dimension which is
discovered in the work presented here.

The issue of dimensionality has become of even greater importance
over the past few decades, with the theories of hidden dimensions
first postulated by Kaluza \cite{kaluza} resurfacing in the quest
for unification and a Theory of Everything. Originally it was hoped
one extra dimension would be sufficient; with the advent of string
theory and its subsequent developments this was then expanded to
twenty six in the late 1960's as a consistency requirement for
bosonic strings, before being reduced back down to ten with the
introduction of supersymmetry in the 1980's. The idea of holography
also has its roots in dimensionality, and its most famous form, that
of Maldacena \cite{adscft}, states a relation between string theory
on a five-dimensional anti-de Sitter space, and a four-dimensional
conformal field theory. Much work (see \cite{aha} for a review) has
followed investigating spacetimes and conformal field theories of
varying dimension and complexity.

Here we present an analysis into the stability of radiating perfect
fluid spheres in an asymptotically anti-de Sitter spacetime. Due to
the confining nature of the AdS potential, the spheres of radiation
are self-gravitating, and thus we shall refer to them as ``stars''
in much of the following, although this is mainly used as a
conveniently brief label, as it is only a toy-model approximation to
a star at best.

We begin by giving the equations for a such a model in
$d$-dimensions, before analysing the behaviour of the star's total
mass. We consider the variation of the total mass as a function of
the central density, and observe that for large enough
dimensionality, the mass increases monotonically with the density.
However in the lower dimensional cases, oscillations appear (this
was originally noted in the $d=5$ case in \cite{hubenynew}),
indicating that the perfect fluid model of the star is becoming
unrealistic. We numerically find the critical dimension separating
these two regimes to be $11.0$ (to three significant figures), and
give an explicit relation, \eqref{eq:ch6peakest1}, between the
spacetime dimension $d$ and the ``saturation density'' $\rho_{c}$,
see section \ref{sec:criticald}. The existence of local maxima at
critical central densities (saturation points) in the lower
dimensional cases indicates the appearance of instabilities in the
model of the star, and point to unrealistic effects developing for
$\rho_{0} > \rho_{c}$. We also provide a numerical analysis of the
behaviour at large central density, in particular the self-similar
behaviour that appears in dimension $d < d_{c}$; several parameters
of our numerical model are then also determined analytically from a
dynamical systems analysis of the behaviour, where we consider the
expansion about a fixed point of the zero-cosmological constant
solution.\footnote{This dynamical systems analysis (section
\ref{sec:dynamsys}) was suggested after correspondence with V.~
Vaganov, who also considered the behaviour of self-gravitating
radiation in $AdS_{d}$ in \cite{vaga}. Work simultaneously conducted
by P.~H.~Chavanis also found the critical dimension described here,
via an alternative route, in his comprehensive study of relativistic
stars with a linear equation of state, \cite{chan} (see the note
following the discussion for more details).}

The outline of the paper is as follows: we begin with a brief recap
on perfect fluid models in general dimension in section
\ref{sec:fluidmodels}. In section \ref{sec:analysis} we introduce
the total mass as a function of the central density, analyse the
progression of the saturation point with increasing dimension, and
present the best fit formula, \eqref{eq:ch6peakest1}, which gives a
value for the critical dimension. We also present further numerical
results for the behaviour of the total mass at large central
densities. In section \ref{sec:dynamsys} we give a dynamical systems
analysis, following the methods in \cite{vaga} and \cite{uggla},
which yields analytical expressions for several of the numerical
results of the previous section. Finally, we conclude with a
discussion of the results in section \ref{sec:discussion}.

\section{Perfect fluid models} \label{sec:fluidmodels}

Consider a general static, spherically symmetric $d$-dimensional AdS
spacetime with metric:
\begin{equation} \label{eq:ch44AdSmetric}
ds^{2} = - k(r) dt^{2} + h(r) dr^{2} + r^{2}d \Omega_{d-2}^{2}
\end{equation}
By considering a perfect fluid of a gas of radiation, one can obtain
implicit expressions for $k(r)$ and $h(r)$ for a simple model of a
``star'' geometry. For a perfect fluid we have that the stress
tensor is of the form:
\begin{equation} \label{eq:ch44stresstensor}
T_{ab} = \rho(r) u_{a} u_{b} + P(r) (g_{a b} + u_{a} u_{b})
\end{equation}
where $u^{a}$ is the $d$-velocity of a co-moving gas, and upon which
we impose the further constraint that the matter be purely
radiating; this sets $\rho(r) = (d - 1) P(r)$ as it requires that
$T_{ab}$ be traceless. One obtains the required metric by solving
Einstein's equation: $G_{a b} + \Lambda g_{a b} = 8 \pi G_{d} T_{a
b}$, with the above stress tensor and a negative cosmological
constant, as follows. The relevant components of Einstein's
equations in general dimension $d$ are given by:

\begin{equation} \label{eq:ch44Grr}
G_{r r} = \left(\frac{d-2}{2}\right) \frac{(d-3)(k(r) - k(r) h(r)) +
r k'(r)}{r^{2} \, k(r)} = \left(\frac{\rho(r)}{d-1} + \frac{(d-1)
(d-2)}{2 \, R^{2}} \right) h(r)
\end{equation}
\begin{equation} \label{eq:ch44Gtt}
G_{t t} = k(r) \left(\frac{d-2}{2}\right) \frac{(d-3)(h^{2}(r) -
h(r)) + r h'(r)}{r^{2} \, h^{2}(r)} = \left(\rho(r) + \frac{(d-1)
(d-2)}{2 \, R^{2}} \right) k(r)
\end{equation}
where we have used that $\Lambda = - (d-1)(d-2)/(2 R^{2})$, $P(r) =
\rho(r)/(d-1)$, and set $8 \pi G_{d} \equiv 1$ for
convenience\footnote{In the numerical results presented shortly we
also set $R = 1$; we include it here to ease comparison with the
dynamical systems analysis of the $\Lambda = 0$ ($R = \infty$) case
given in section \ref{sec:dynamsys}.}. We can infer the form of
$h(r)$ from \eqref{eq:ch44Gtt}, as the $k(r)$ dependence cancels,
and we find that $h(r)$ is given by:
\begin{equation} \label{eq:ch6starhr}
h(r) = \left(1 + \frac{r^{2}}{R^{2}} -
\frac{m(r)}{r^{d-3}}\right)^{-1}
\end{equation}
where $m(r)$ is a mass function\footnote{Note that the mass function
used here is a rescaling of the actual mass; a constant factor from
the integral over the angular directions does not appear in our
definition of $m(r)$ due to our definition of $h(r)$.} related to
the density via:
\begin{equation} \label{eq:ch6starmr}
m(r) = \frac{2}{d - 2} \int_{0}^{r} \rho(\acute{r}) \acute{r}^{d-2}
\, \mathrm{d} \acute{r}
\end{equation}

In order to specify a form for $k(r)$, we recall the energy-momentum
conservation equation, $\nabla_{\mu} T^{\mu \nu} = 0$, which for a
general perfect fluid without the radiation condition gives:
\begin{equation} \label{eq:ch6energymtmcons}
P'(r) + \frac{k'(r)}{2 \, k(r)} (P(r) + \rho(r)) = 0
\end{equation}
which can be re-arranged to give
\begin{equation} \label{eq:ch6starfr}
k(r) = \left(\frac{\rho_{\infty}}{\rho(r)} \right)^{2/d}
\end{equation}
in the radiation case, where we have introduced $\rho_{\infty}$,
which is the leading coefficient of $\rho(r)$ at large $r$, and is
given by $\rho_{\infty} \approx \rho(r) r^{d}$ as $r \rightarrow
\infty$. Substituting $h(r)$ from \eqref{eq:ch6starhr} into
\eqref{eq:ch44Grr} and eliminating $k'(r)/k(r)$ using
\eqref{eq:ch6energymtmcons} then gives an equation in terms of
$m(r)$, $\rho(r)$ and $\rho'(r)$,.
\begin{equation} \label{eq:ch6starode1}
\frac{(d - 3)}{r^{2}} \left(1 - \frac{1}{1 + \frac{r^{2}}{R^{2}} -
\frac{m(r)}{r^{d-3}}}\right) - \frac{2 \rho'(r)}{r \, \rho(r) \, d}
= \frac{2 \rho(r) + (d-1)^{2}(d-2)/R^{2}}{(d -1)(d-2) \left(1 +
\frac{r^{2}}{R^{2}} - \frac{m(r)}{r^{d-3}}\right)}
\end{equation}
which couples with our equation for $m'(r)$:
\begin{equation} \label{eq:ch6starode2}
m'(r) = \frac{2}{d-2} \, \rho(r) \, r^{d-2}
\end{equation}
to give a pair of ODEs. For specified dimension $d$, these allow the
geometry of the spacetime to be numerically generated when they are
combined with the relevant boundary conditions: $m(0) = 0$ and
$\rho(0) = \rho_{0}$. The condition $\rho(0) = \rho_{0}$ specifies
the central density of the gas, and we have that for fixed $R$,
$\rho_{0}$ is the single free parameter of the system (pure AdS is
recovered when $\rho_{0} = 0$).

\section{Total mass as a function of central density}
\label{sec:analysis}

Whilst mathematically one can work with this perfect fluid setup in
any number of dimensions (including non-integer ones), one would
also like to consider the appropriateness of doing so, given that we
wish to use the geometry as the setup for a toy model of a star. In
other words, is there any significant change in behaviour as the
dimensionality of the model is altered. A particular quantity of
interest in analysing the stability of the model is the total mass
$M$ of the star, and as we have just seen, the mass and density
profiles of our gas of radiation are determined by a single
parameter: the central density of the gas, $\rho_{0}$.

To avoid possible instabilities such as those considered in the
asymptotically flat case (in four dimensions) in \cite{sorkin}, one
would expect the total mass to increase monotonically with
$\rho_{0}$. One could also expect the total mass to be bounded from
above by some maximum value, analogous to the $4 d$ asymptotically
flat case where for a fixed size $R_{star}$, the maximum possible
mass such a star can have is given by $M_{max} = 4 R_{star}/9$, a
result found by Buchdahl in 1959 \cite{buch}.\footnote{For further
edification, note that this equality coincides with the
Buchdahl-Bondi limit, usually written in the form $R/M = 9/4$, which
is the lowest radius to which the Schwarzschild geometry can be
embedded (in Euclidean space). For more detail see e.g.
\cite{heinznew,abramnew}.}

\begin{figure}
\begin{center}
  \includegraphics[width=0.9\textwidth]{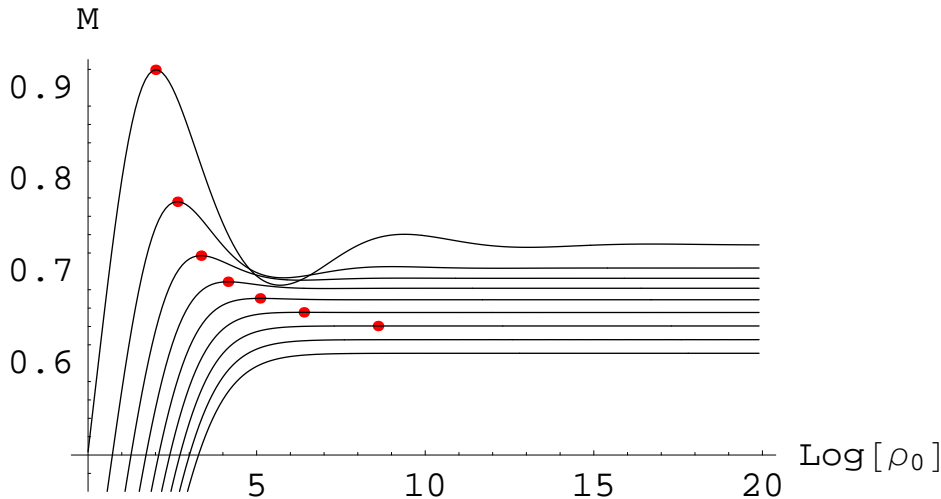}\\
\end{center}
\vspace{-1.0cm} \caption{Total mass vs density for the radiating
perfect fluid model in various dimensions, from $d=4$ (top curve)
through to $d = 12$ (bottom curve). The saturation point for each
dimension is indicated by the red dots; these correspond to the
maximum value of the total mass in the relevant dimension, at the
critical density $\rho_{c}$ (see section \ref{sec:criticald}). For
$d$ large, there is no local maximum and hence no finite saturation
point; in these cases, the maximum total mass is given by the
asymptotic value, $\eta_{d}$. }\label{sat1}
\end{figure}

\begin{figure}
\begin{center}
  \includegraphics[width=0.9\textwidth]{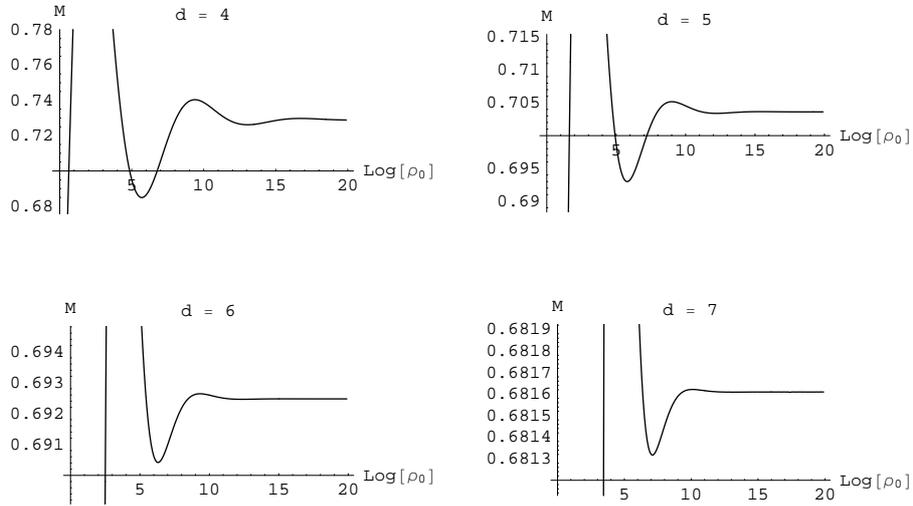}\\
\end{center}
\vspace{-1.0cm} \caption{The oscillations in the total mass $M$: as
the central density $\rho_{0}$ is increased, $M$ does not simply
increase monotonically towards its final value $\eta_{d}$. Instead,
it reaches a larger maximum before undergoing damped oscillations
towards $\eta_{d}$. Note that the amplitude of the oscillations
becomes smaller as the dimensionality $d$ is
increased.}\label{lowdplot}
\end{figure}

\begin{figure}
\begin{center}
  \includegraphics[width=0.9\textwidth]{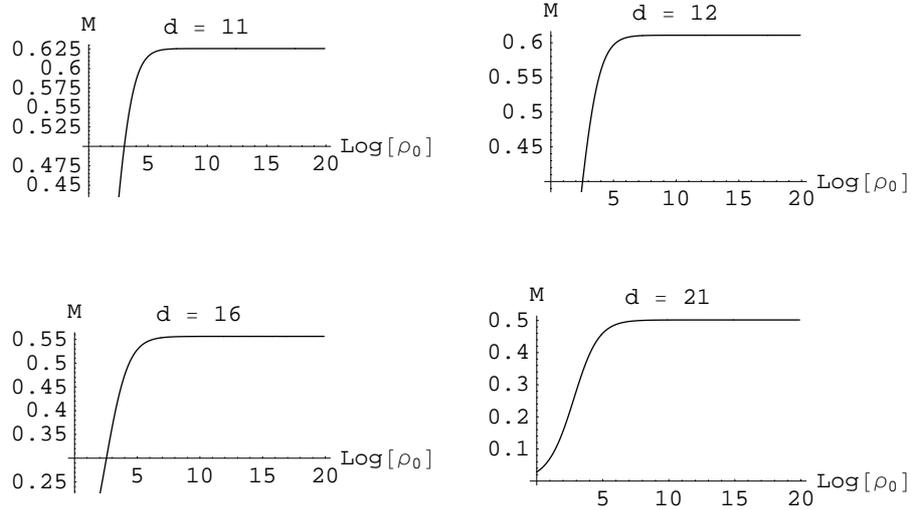}\\
\end{center}
\vspace{-1.0cm} \caption{For larger $d$, there are no oscillations
in the total mass: $M$ is now a monotonic function of the central
density $\rho_{0}$, and its maximum is also its asymptotic value as
$\rho_{0} \rightarrow \infty$, namely $\eta_{d}$.}\label{highdplot}
\end{figure}

In our scenarios the total mass is indeed bounded from above,
however, this maximum is not always the asymptotic value of the
total mass at large density. Although we observe that as $\rho_{0}
\rightarrow \infty$ we have $M(\rho_{0}) \rightarrow \eta_{d}$,
where $\eta_{d}$ is some finite constant dependent on the dimension
$d$ (see section \ref{sec:masslargerho} below for more details),
what we do not find in all cases is the total mass approaching this
constant monotonically, see figure \ref{sat1}. When the
dimensionality is low, there are sizable oscillations about the
final value $\eta_{d}$ before the curve settles down (see figure
\ref{lowdplot}), as was noted in the $d=5$ case in \cite{hubenynew},
and in other similar scenarios, e.g. \cite{donp}, and the star's
maximum mass is given by some value greater than $\eta_{d}$. As the
dimension is increased, however, these oscillations become less
pronounced, and for $d$ sufficiently high they disappear altogether,
see figure \ref{highdplot}.

Ideally one would like to analytically determine the dependence of
the shape of the curve on both the dimension $d$ and the density
$\rho_{0}$, however, due to the complexity of the equations, the
exact behaviour must be computed numerically. One can nevertheless
use this data to construct models of the various features of the
star's behaviour: for example, in section \ref{sec:masslargerho}
below, we give an analysis of the total mass at large $\rho_{0}$
(where it approaches a constant, dependent on $d$) in different
dimensions.

One particularly interesting feature is the appearance of the
turning points in the total mass seen in figure \ref{lowdplot}, and
specifically the locations of the local maxima in different
dimensions. One can see from the figures that as the dimensionality
is increased, the appearance of the first maximum moves to larger
$\rho_{0}$; by analysing this progression one can obtain a
remarkably simple relation which immediately gives a value for the
critical dimension, above which the oscillations do not exist, and
hence the total mass is a monotonic function of $\rho_{0}$.

\subsection{A critical dimension} \label{sec:criticald}

\begin{figure}
\begin{center}
  \includegraphics[width=0.9\textwidth]{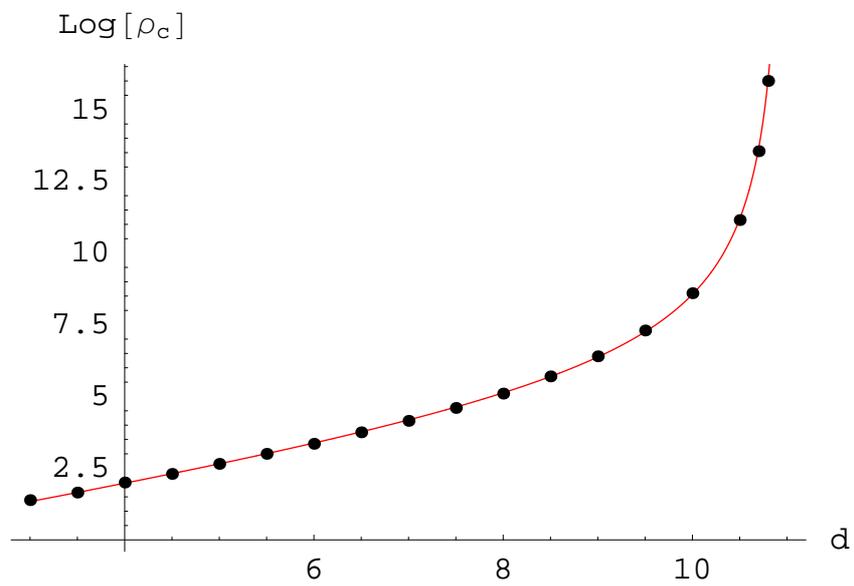}\\
\end{center}
\caption{The change in the saturation point $\rho_{c}$ with
increasing dimension $d$. The points plotted are the calculated
values for the saturation point for the star model in the
corresponding dimension, the red best fit line is the curve given by
\eqref{eq:ch6peakest1}. The divergent behaviour as $d$ approaches
eleven indicates that for $d > 11$ there is no saturation point, and
hence no apparent instability in the perfect fluid model of the
``star''. }\label{peak1}
\end{figure}

The saturation point, $\rho_{c}$, which we define as being the
location of the first local maximum when increasing $\rho_{0}$, can
be seen to progress to larger and larger $\rho_{0}$ as the dimension
$d$ is increased, see figure \ref{sat1}. What we wish to determine
is whether this saturation point appears for all dimension $d$ (for
sufficiently large $\rho_{0}$), or whether there is a cut-off
dimension, $d_{c}$, such that for larger $d$, there is no local
maximum and hence no saturation point. Figure \ref{peak1} shows how
the saturation point varies with dimension; numerical analysis then
finds (to 3 significant figures) that this behaviour is given by the
following model:

\begin{equation} \label{eq:ch6peakest1}
\log{\rho_{c}} \approx 0.500 \, d + \frac{5.75}{\sqrt{11.0 - d}} -
2.20
\end{equation}
which gives a critical dimension $d_{c} = 11.0$.\footnote{As
mentioned earlier, correspondence with V.~Vaganov and P.~H.~Chavanis
suggested that the critical dimension in the radiating perfect fluid
case is very close to (but not exactly) eleven, and this is indeed
the case as we see in the dynamical systems analysis approach in
section \ref{sec:dynamsys}, where we obtain a value complementary to
the numerical estimate of $11.0$ given here. Interestingly, the
exact value of $d_{c} = 11$ appears in the case of Newtonian
isothermal spheres, as noticed by Sire and Chavanis in 2002
\cite{chan2}.} What is perhaps rather surprising is the simplicity
of \eqref{eq:ch6peakest1}: not only do we have a critical dimension
appearing so clearly, the overall dependence on $d$ is remarkably
simple, and the co-efficient of the linear term appears to be
exactly one half.

As we shall see in section \ref{sec:dynamsys}, the value of the
critical dimension can also be determined by an analytical
consideration of the radiating perfect fluid system with zero
cosmological constant (i.e. in the limit $R \rightarrow \infty$).
Although such a solution is singular at $r = 0$, and has infinite
mass, by confining the radiation to finite sized box one can obtain
finite mass solutions. The features determined in this configuration
can be related to equivalent behaviour in the asymptotically anti-de
Sitter case (where the (finite) mass is confined by the AdS
potential), and indeed exact values for certain parameters can also
be computed. This same analysis is not restricted to the star
geometries considered here, it can be used with any linear equation
of state \cite{vaga}, or even more generally \cite{uggla}. Before
giving the analysis for our case of perfect fluid radiation,
however, we firstly present further numerical results.

\subsection{Total mass at large $\rho_{0}$} \label{sec:masslargerho}

In addition to considering the variation of the saturation point for
the star with dimension, one can also investigate the asymptotic
behaviour of $M$ as $\rho_{0}$ becomes large. As mentioned in
section \ref{sec:analysis}, at large $\rho_{0}$, the value of the
total mass tends to a constant, $\eta_{d}$, which is then only
dependent on the dimension; the value of this constant decreases as
$d$ increases. The values are plotted in figure \ref{etaplot1} and
despite the complicated appearance of the plot, a remarkably close
fit for all dimensions is given by:

\begin{figure}
\begin{center}
  \includegraphics[width=0.45\textwidth]{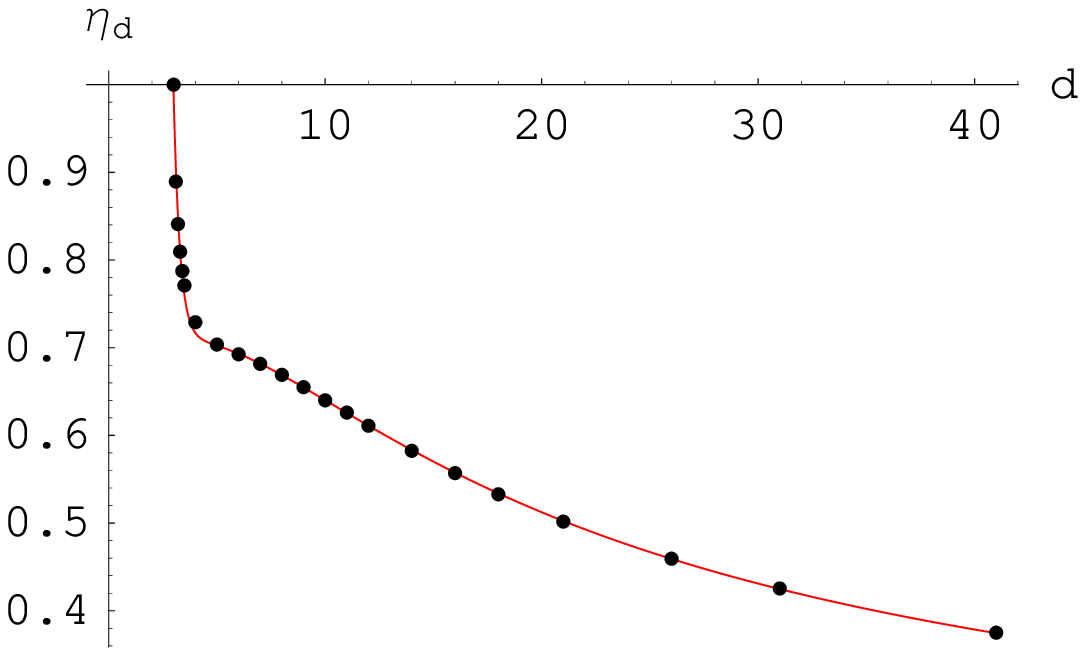}
  \includegraphics[width=0.45\textwidth]{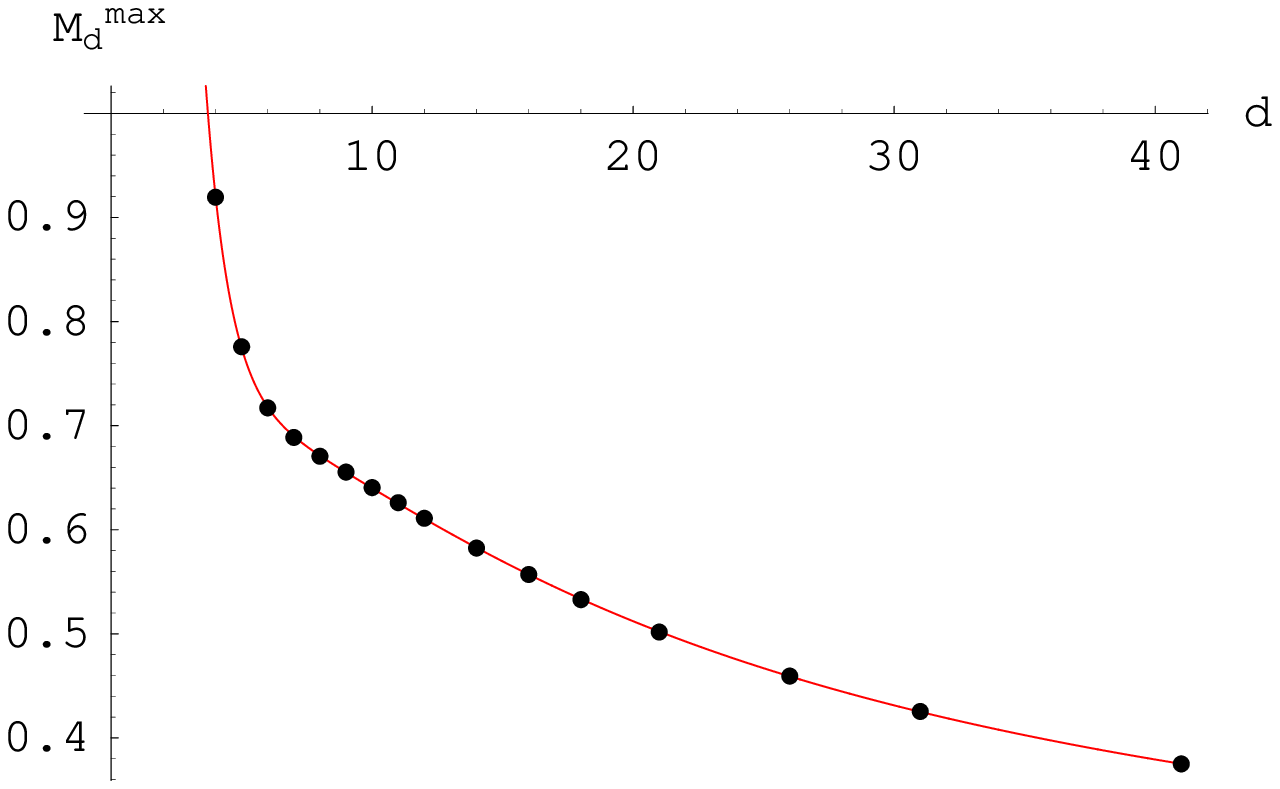}\\
\end{center}
\caption{The plot on the left shows $\eta_{d}$ for various
dimensions, with the approximation given in \eqref{eq:eta1} shown in
red. The data points are all at integer values for the dimension,
with the addition of points at $d = 3.1, 3.2,...,3.5$ to highlight
the behaviour of the curve at low $d$. The righthand plot shows the
behaviour of $M_{d}^{max}$; this is identical to that of $\eta_{d}$
for $d \ge 11$, however for $d < 11$, the maximum is given by the
value of the total mass at the saturation point, $\rho_{c}$. The
best fit approximation (red curve) for each is simple in terms of
its $d$ dependence, and provides a good fit over a large range of
$d$.}\label{etaplot1}
\end{figure}

\begin{equation} \label{eq:eta1}
\eta_{d} \approx 0.716 + \exp \left[9.85 - 3.72 \, d \right] - \exp
\left[- 0.603 - \frac{20.3}{d} \right]
\end{equation}
which is also shown in the figure. Checks show that the function
continues to give accurate predictions for larger $d$, and although
there is perhaps slightly more complicated behaviour for $d \sim 4$,
we do not attempt to investigate this further here\footnote{Whilst
in the dynamical systems analysis (section \ref{sec:dynamsys}) the
$d = 3$ case (where we have $\eta_{3} = 1$) needs considering
separately, as there is different asymptotic behaviour involved due
to the non-dynamical nature of gravity in such a scenario (see
\cite{hammer2} for example), we find that we can include it both
here (in the analysis of $\eta_{d}$) and also in our earlier result
for the critical dimension (see section \ref{sec:criticald}).};
despite the relative compactness of the expression, there is little
intuitive origin for any of the constants involved. Nonetheless, it
is impressive that the behaviour of the mass at large $\rho_{0}$ can
be expressed in such simple powers of the dimension.

One can perform a similar analysis of the behaviour of the maximum
value of the total mass as the dimension increases; the results are
also shown in figure \ref{etaplot1}. For $d > 11$, the maximum total
mass corresponds to the asymptotic value, $\eta_{d}$, however, for
lower dimension, the maximum is given by the mass at the saturation
point. A good fit to the curve is given by:
\begin{equation} \label{eq:eta2}
M_{d}^{max} \approx 0.712 + \exp \left[2.74 - 1.07 \, d \right] -
\exp \left[- 0.592 - \frac{20.5}{d} \right]
\end{equation}
which differs (significantly) from \eqref{eq:eta1} only in the
second term; this was to be expected, as the curves differ only at
low $d$. Whilst their is little apparent significance about the
values of the numerical constants involved in the expression, we
have again produced a fit with a relatively simple dependence on $d$
which gives accurate predictions for $M_{d}^{max}$ over a large
range of dimensions.

Although the form of the fit used in equations \eqref{eq:eta1} and
\eqref{eq:eta2} was chosen primarily because it gave such a close
fit to the data, it would be interesting to study possible reasons
for expecting the observed dependence on dimension, although we do
not pursue this further here. What we do now examine in more detail,
is the oscillatory behaviour, which can be considered both
numerically and analytically.

\subsection{Self-similarity analysis for $d < 11$}
\label{sec:selfsim}

Another interesting feature of the plots of the total mass seen in
figures \ref{sat1} and \ref{lowdplot} is the self-similarity
exhibited by the oscillatory behaviour as $\rho_{0} \rightarrow
\infty$. A numerical analysis of the periodicity and damping of the
oscillations seen for $3 < d < 11$ leads us to propose the following
model for the total mass:
\begin{equation} \label{eq:osc1}
M_{d} (\rho_{0}) \approx \eta_{d} + \alpha_{d} \exp \left[-
\beta_{d} \, \log(\rho_{0}) \right] \cos\left[\mu_{d} - \nu_{d}
\log(\rho_{0})\right]
\end{equation}
which gives a good approximation for the behaviour in the region
$\rho_{0} > \rho_{c}$. In \eqref{eq:osc1}, $\eta_{d}$ is the
asymptotic value of the mass discussed above, and the four
parameters $\alpha_{d}$, $\beta_{d}$, $\mu_{d}$ and $\nu_{d}$ are
constants dependent only on the dimension $d$. Approximate values
for these constants for $d = 3.1, 4, 5, 6$ and $7$ are given in
Table \ref{addDtab1}.

\begin{table}
\begin{center}
\begin{tabular}{l l l l l}
\hline $d$ & $\alpha_{d}$ & $\beta_{d}$ & $\mu_{d}$ & $\nu_{d}$\\
\hline 3.1 & 0.305 & 0.184 & 8.33 & 0.66 \\
4 & 0.383 & 0.371 & 8.44 & 0.86 \\
5 & 0.400 & 0.601 & 9.35 & 0.98 \\
6 & 0.415 & 0.825 & 10.3 & 1.03 \\
7 & 0.431 & 1.03 & 11.4 & 1.07 \\
\hline
\end{tabular}
\end{center}
\caption{Numerical estimates for $\alpha_{d}$, $\beta_{d}$,
$\mu_{d}$ and $\nu_{d}$ (to three significant figures) for the model
of the total mass given in \eqref{eq:osc1} .}\label{addDtab1}
\end{table}

Although the values given in table \ref{addDtab1} are only
approximate, we nonetheless see interesting dependencies on $d$
emerging. For example, $\beta_{d}$ appears to increase roughly
linearly with dimension ($\beta_{d} \approx 0.22 d - 0.51$), as do
$\alpha_{d}$ and $\mu_{d}$ for $d \ge 4$. We will see below in the
dynamical systems analysis how this linear behaviour of $\beta_{d}$
on $d$ is only an approximation to the true behaviour, and the same
analysis also provides exact values for the parameter $\nu_{d}$.
This analytical analysis also confirms the form of the fit used in
the numerical approximation derived above.

\section{Dynamical systems analysis} \label{sec:dynamsys}

By considering the behaviour of the system of coupled ODEs given in
section \ref{sec:fluidmodels} in the limit $R \rightarrow \infty$,
we can obtain analytical results for some of the interesting
features of the radiating perfect fluid star geometries described
above. The analysis presented here follows that detailed in both
\cite{vaga} and \cite{uggla}, where it is given in more general
settings; by focusing on the radiation case (where $\rho(r) = (d-1)
P(r)$) we can give a good explanation of why the numerical behaviour
seen above is so, without excessive over-complication.

The basic idea is to rewrite the equations for $\rho'(r)$ and
$m'(r)$ in terms of dimensionless (compact) variables and perform an
analysis of the fixed points. The corresponding eigenvalues and
eigenvectors obtained by linearising about these fixed points give a
complete description of the nearby behaviour (Hartman-Grobman
theorem, \cite{hart}) on the new state space, which can then be
translated back to the physical picture by inverting the
transformations given below. Interestingly, for the perfect fluid
stars, the dependence of the total mass (as well as other
quantities, e.g. the entropy) on the central density, $\rho_{0}$, is
governed by the behaviour around (and hence the eigenvalues of) a
single fixed point. Specifically for our case we will see how this
gives both an exact value for the critical dimension $d_{c}$, and a
clear analytical explanation for the observed behaviour in the two
regimes $d < d_{c}$ (oscillatory) and $d > d_{c}$ (monotonically
increasing). We will also obtain expressions for the $\beta_{d}$ and
$\nu_{d}$ parameters introduced earlier.

To proceed, we thus set $R = \infty$, and our equations
\eqref{eq:ch6starode1} and \eqref{eq:ch6starode2} become:
\begin{equation} \label{eq:ch6dynam1}
\rho'(r) = - \frac{\rho(r) \, d \left( (d-3)(d-2)(d-1) m(r) + 2
r^{d-1} \rho(r) \right)}{2 (d-2)(d-1)(r^{d-2} - r \, m(r))}
\end{equation}
\begin{equation} \label{eq:ch6dynam2}
m'(r) = \frac{2}{d-2} \, \rho(r) \, r^{d-2}
\end{equation}
where $8 \pi G_{d}$ has again been set equal to one. Note that we do
not include the $d = 3$ scenario here as it is a special case (due
to the non-dynamical nature of gravity). We can now introduce the
dimensionless variables:
\begin{equation} \label{eq:ch6dynam3}
u = \frac{2 \, r^{d-1} \rho(r)}{(d-2) m(r)}
\end{equation}
and
\begin{equation} \label{eq:ch6dynam4}
v = \frac{(d-1) \, m(r)}{2 \, r^{d-3}} \left(1 -
\frac{m(r)}{r^{d-3}} \right)^{-1}
\end{equation}
which allow equations \eqref{eq:ch6dynam1} and \eqref{eq:ch6dynam2}
to be rewritten in the form:
\begin{equation} \label{eq:ch6dynam5}
\frac{d u}{d \xi} = - u \left(1 - d + u + \left(d - 3 +
\frac{u}{d-1}\right)\left(\frac{v \, d}{d-1}\right)\right)
\end{equation}
\begin{equation} \label{eq:ch6dynam6}
\frac{d v}{d \xi} = - v \left(d - 3 - u + \left(d - 3 - u \right)
\left(\frac{2 \, v}{d-1}\right)\right)
\end{equation}
where we have also introduced the new independent variable $\xi =
\ln(r)$. For the case of positive mass and density we're considering
here, both $u$ and $v$ are greater than zero (for non-zero $r$), and
we make a final change of variables to the bounded $U$ and $V$
defined by\footnote{Although the range of both $U$ and $V$ is
defined as being $(0,1)$, in order to perform the fixed point
analysis of the asymptotic behaviour, it is necessary that the
boundary points also be included; this requires the system given by
\eqref{eq:ch6dynam8} and \eqref{eq:ch6dynam9} be $\mathcal{C}^{1}$
on $[0,1]^{2}$, which is manifestly so.}:
\begin{equation} \label{eq:ch6dynam7}
U = \frac{u}{1+u} \, , \;\;\;\;\;\; V = \frac{v}{1+v}
\end{equation}
which gives the system of equations:
\begin{equation} \label{eq:ch6dynam8}
\frac{d U}{d \lambda} = U (1-U) \left( d-1- d U - \left(2 d -4 +
\frac{d-3}{d-1} \right) V + \left(2 d - 3 + \frac{d(d-5)
+3}{(d-1)^{2}}\right) U V \right)
\end{equation}
\begin{equation} \label{eq:ch6dynam9}
\frac{d V}{d \lambda} = V (1-V)(3 - d + (d-2)
U)\left(1+\left(\frac{2}{d-1} - 1\right) V\right)
\end{equation}
where we have also introduced the independent variable $\lambda$,
defined by:
\begin{equation} \label{eq:ch6dynam10}
\frac{d \lambda}{d \xi} = \frac{1}{(1-U)(1-V)} \;\;\;\;\; \Bigg( =
(1+u)(1+v) \Bigg)
\end{equation}

The fixed points of the system (\eqref{eq:ch6dynam8} and
\eqref{eq:ch6dynam9}) are calculated in the usual fashion, by
setting both $d U/d \lambda$ and $d V/d \lambda$ to zero and solving
for $U$ and $V$; there are six in total, with eigenvalues then
obtained from
\begin{equation} \label{eq:ch6dynam11}
\frac{d}{d \lambda}\left( \begin{array}{c} U \\
V \end{array} \right) = \left( \begin{array}{cc}
\frac{\partial}{\partial U} \left(\frac{d U}{d \lambda}\right) & \frac{\partial}{\partial V} \left(\frac{d U}{d \lambda}\right)\\
\frac{\partial}{\partial U} \left(\frac{d V}{d \lambda}\right) &
\frac{\partial}{\partial V} \left(\frac{d V}{d \lambda}\right)
\end{array}
\right)\Bigg|_{fp} \left( \begin{array}{c} U - U_{fp}\\
V - V_{fp}\end{array} \right)
\end{equation}
where the matrix components are evaluated at the particular fixed
point under consideration. A table of such eigenvalues is given in
\cite{vaga}, where they are labelled $T_{1}, \dots, T_{6}$; we do
not list them all again here, however, as orbits in the interior of
the state space $[0,1]^{2}$ originate from either $T_{2}$ or $T_{4}$
and converge to the fixed point $T_{3}$ (as is shown in
\cite{uggla}). This fixed point corresponds to the singular
self-similar solution given by equation (2.14) of \cite{vaga}, and
due to the scale invariance of the system one can consider the
entire set of (positive mass) solutions as being represented by a
single orbit from $T_{2}$ to $T_{3}$ (this is true for any linear
equation of state, $P(r) = q \rho(r)$).

Although one cannot write an analytic expression for this orbit, one
can obtain approximations by linearising about the fixed points. As
discussed briefly earlier, as this zero-cosmological constant
solution is singular, in order to produce finite mass solutions the
radiation must be confined to an (unphysical) box; the two fixed
points $T_{2}$ and $T_{3}$ thus represent solutions with $\rho_{0} =
0$ and in the limit $\rho_{0} \rightarrow \infty$ respectively. The
behaviour described by the linearisation about $T_{3}$ then reveals
aspects of the large $\rho_{0}$ limit of the radiating stars (where
the confining AdS potential results in the finite mass solutions
without the need for any unphysical box), exactly what we analysed
numerically in sections \ref{sec:masslargerho} and
\ref{sec:selfsim}. This linearisation gives an explanation for the
existence of a critical dimension and the differing behaviour seen
in higher and lower dimensions, including quantitative expressions
for $d_{c}$ and the $\beta_{d}$ and $\nu_{d}$ parameters of Table
\ref{addDtab1}, as we shall now show.

Fixed point $T_{3}$ corresponds to the following values of $U$ and
$V$:
\begin{equation} \label{eq:ch6dynam12}
U_{T_{3}} = \frac{d-3}{d-2} \, , \;\;\;\;\;\; V_{T_{3}} = \frac{2
(d-1)^{2}}{2 - 4 \, d + (d-1)d^{2}}
\end{equation}
and has eigenvalues:
\begin{eqnarray} \label{eq:ch6dynam13}
T_{3}^{\pm} && \hspace{-0.5cm} = \frac{d (d-3)}{2- 4 \, d +
(d-1)d^{2}}\left( 1-d \pm \sqrt{\frac{(d-12)d^{2} + 13 \, d -
18}{d-2}} \right)  \nonumber \\
&& \hspace{-0.5cm} \equiv k_{d} \left( 1-d \pm
\sqrt{\frac{(d-12)d^{2} + 13 \, d - 18}{d-2}} \right)
\end{eqnarray}
where we denote the coefficient $k_{d}$ and observe that it is
strictly positive for $d > 3$. These eigenvalues govern the
behaviour of the solution, and we immediately see that there are two
distinct regimes; one where the expression inside the square root is
negative, corresponding to the oscillatory behaviour seen in figure
\ref{lowdplot}, and one where the expression is positive, resulting
in the monotonic behaviour seen in figure
\ref{highdplot}.\footnote{The fact that $k_{d} > 0$ ensures that the
fixed point $T_{3}$ is a stable focus for the oscillatory behaviour
in the $d < d_{c}$ case; for $d > d_{c}$, we have that $T_{3}^{\pm}$
is strictly less than zero, and hence acts as a stable node.} We
thus obtain a value for the critical dimension given by the solution
to:
\begin{equation} \label{eq:ch6dynam14}
(d_{c}-12)d_{c}^{2} + 13 \, d_{c} - 18 = 0
\end{equation}
which yields $d_{c} = 10.964\dots$, complementary to the value of
$d_{c} = 11.0$ obtained numerically, although with the significance
of being non-integer rather than exactly $11$; interestingly for any
linear equation of state the value of $d_{c}$ is always in the range
$10 \le d_{c} \le 11$, see \cite{vaga}.

We can relate the asymptotic behaviour obtained from the state space
picture to the physical quantities of mass and density via several
auxiliary equations to those given above, specifically:
\begin{equation} \label{eq:ch6dynam15}
\frac{d r}{d \lambda} = (1 - U)(1 - V) r \, , \;\;\; \frac{d m}{d
\lambda} = U(1 - V) m
\end{equation}
and
\begin{equation} \label{eq:ch6dynam16}
\frac{d \rho}{d \lambda} = - \frac{V \, d}{d-1}\left(1 - U +
\frac{U}{d-1} \right) \rho
\end{equation}

Given expressions for $U$ and $V$ in terms of $\lambda$ (as obtained
from an analysis of the behaviour around the fixed points, say), one
can integrate the above to determine corresponding expressions for
the mass, radius and density in terms of $\lambda$. There is,
however, a simple way to see the dependence of the total mass
$M_{d}$ on the central density $\rho_{0}$, which also reveals the
origin of the $\beta_{d}$ and $\nu_{d}$ parameters of our numerical
model in section \ref{sec:selfsim}.

Focusing then on the case where $d < d_{c}$, how does the imaginary
term in \eqref{eq:ch6dynam13} lead to the (self-similar) oscillatory
behaviour manifest in the total mass at large $\rho_{0}$? This can
be seen directly from the linearisation about $T_{3}$, where we
observe similar oscillations in the expressions for $U(\lambda)$ and
$V(\lambda)$ (see below); as mentioned above, it is the behaviour
around $T_{3}$ that governs the behaviour of the physical quantities
in the large $\rho_{0}$ limit. By considering the behaviour of
$U(\lambda)$ and $V(\lambda)$ in terms of $\rho_{0}$, we can extract
the coefficients which should then match those in \eqref{eq:osc1}
(as argued more fully in \cite{uggla}). The solutions of
\eqref{eq:ch6dynam8} and \eqref{eq:ch6dynam9} in the large $\lambda$
limit (i.e. about the fixed point $T_{3}$) can be expressed as:
\begin{equation} \label{eq:ch6dynam15b}
Re(U(\lambda)) = U_{T_{3}} + \exp\left(- (d-1) k_{d} \lambda\right)
\cos\left(k_{d} \lambda \sqrt{\frac{18-(d-12)d^{2} - 13 \, d}{d-2}}
\right)
\end{equation}
\begin{equation} \label{eq:ch6dynam16b}
Re(V(\lambda)) = V_{T_{3}} + \exp\left(- (d-1) k_{d} \lambda\right)
\cos\left(k_{d} \lambda \sqrt{\frac{18 -(d-12)d^{2} - 13 \, d}{d-2}}
\right)
\end{equation}
where we have only kept the real angular term as we are only
interested in the period of the oscillations ($\nu_{d}$) and the
coefficient of the damping ($\beta_{d}$); the extra factors (namely
$\alpha_{d}$ and $\mu_{d}$) cannot be extracted directly from this
analysis.\footnote{Technically, for a solution of this form one
should first make a linear change of coordinates such that the
matrix on the RHS of \eqref{eq:ch6dynam11} is in diagonal form. This
only manifests itself, however, as extra multiplicative constants
which do not affect the decay term $\beta_{d}$ or oscillation period
$\nu_{d}$, and hence can be ignored.}

For sufficiently high density stars (i.e. with large $\rho_{0}$), we
have $\lambda \propto \frac{1}{2} \frac{d \lambda}{d
\xi}\big|_{T_{3}} \log(\rho_{0})$, and we thus obtain:
\begin{equation} \label{eq:ch6dynam17}
\beta_{d} = \frac{(d-1) k_{d}}{2 (1- U_{T_{3}})(1- V_{T_{3}})} =
\frac{d}{4} + \frac{1}{2 \, d} - \frac{3}{4}
\end{equation}
and
\begin{eqnarray} \label{eq:ch6dynam18}
\nu_{d} && \hspace{-0.5cm} = \frac{k_{d}}{2 (1- U_{T_{3}})(1-
V_{T_{3}})} \sqrt{\frac{18 -(d-12)d^{2} - 13 \, d}{d-2}}  \nonumber \\
&& \hspace{-0.5cm} = \frac{1}{4 \, d} \sqrt{(d-2)(18 -(d-12)d^{2} -
13 \, d)}
\end{eqnarray}
which give the values shown in Table \ref{addDtab1exact}, provided
as a comparison to the numerical estimates obtained in section
\ref{sec:selfsim}. We see that they match very closely, with any
discrepancies most likely due to a combination of numerical
imprecision in the original data for the mass at large $\rho_{0}$
and the use of oscillations at insufficiently large $\rho_{0}$ for
the asymptotic dependence to be totally dominant.

\begin{table}
\begin{center}
\begin{tabular}{l l l}
\hline $d$ & $\beta_{d}$ & $\nu_{d}$\\
\hline 3.1 & $231/1240 \approx 0.186$ & $\sqrt{695519}/1240 \approx 0.672$\\
4 & $3/8 = 0.375$ & $\sqrt{47}/8 \approx 0.857$\\
5 & $3/5 = 0.6$ & $2 \sqrt{6}/5 \approx 0.980$ \\
6 & $5/6 \approx 0.833$ & $\sqrt{13/3}/2 \approx 1.04$\\
7 & $15/14 \approx 1.07$  & $2 \sqrt{215}/5 \approx 1.05$ \\
\hline
\end{tabular}
\end{center}
\caption{Exact values (alongside decimal equivalents) obtained from
the dynamical systems analysis for $\beta_{d}$ and $\nu_{d}$ for the
model of the total mass \eqref{eq:osc1}.}\label{addDtab1exact}
\end{table}

\section{Discussion} \label{sec:discussion}

What we have seen in the above analysis is firstly the appearance of
a critical dimension ($d_{c} = 11.0$) from the simple requirement of
a stability condition on our perfect fluid model for a gas of
radiation, and a numerical study of the behaviour in various
dimensions. Whilst the appearance of oscillations in the variation
of the total mass with the central density $\rho_{0}$ in the lower
dimensional cases had been noted before, we saw here that such
oscillations do not appear to persist in the higher dimensional
cases (see figures \ref{lowdplot} and \ref{highdplot}). Not only do
the oscillations die down as the dimensionality is increased, by
analysing the progression of the saturation point, $\rho_{c}$, we
find that they disappear completely for $d$ above a certain value.
Although this value was calculated to be $11.0$ in the original
numerical analysis, the dynamical systems approach which followed in
section \ref{sec:dynamsys} not only gave a more precise, non-integer
value ($d_{c} = 10.964\dots$, as first derived in \cite{vaga}), but
also explained analytically why one sees a change from oscillatory
to monotonic behaviour as one increases the dimension past this
point. The section concluded by demonstrating how the $\beta_{d}$
and $\nu_{d}$ parameters of the numerical model could also be
derived via this analytical approach.

What is remarkable is that from a seemingly basic condition (that of
monotonicity in the variation of the total mass), one arrives at
such a simple relation for the dependence of $\rho_{c}$ on $d$,
namely equation \eqref{eq:ch6peakest1}. Such simplicity could not
have been expected given the complex nature of the initial setup,
which allows the spacetimes in question to be generated only
numerically from the coupled ODEs. The extensions
\cite{dema1,dema2,elsk} to the BKL work on modelling a gravitational
field close to a spacelike singularity also reveal a critical
dimension of eleven.\footnote{Briefly, their analysis of the setup
was performed using the mixmaster model, where the dynamical
behaviour is governed by Kasner exponents and conditions upon them,
and in which the evolution continues until the system reaches a
stability region where the Kasner exponents remain constant. They
observed that such a stability region could only exist for $d \ge
11$, and thus the evolution continues indefinitely for any lower
number of dimensions. This has interesting consequences not
elaborated on here, which are discussed in detail in the papers
cited above.} In their work they found that the general behaviour of
the relevant Einstein solutions changed from ``chaotic'' in the low
dimensional cases ($d < 11$) to non-chaotic in higher dimensions ($d
\ge 11$), in much the same manner as we observe the transition from
oscillatory to monotonic total mass behaviour in the radiating star
case considered here. It is interesting that their work also reveals
a critical dimension of eleven, and a more detailed comparison of
the two different scenarios (including an analysis to determine the
exact (possibly non-integer) value of $d_{c}$ for their transition)
may yield further insight.

It would be interesting to see if such a result appears in other
investigations into scenarios similar to the radiating perfect fluid
model considered here. For example, one could examine other physical
equations of state to see if they exhibit the same behaviour, and
indeed the work of \cite{vaga} and \cite{chan} has pursued this idea
further (see the note below). Finally, one could also look to
explain the linear $d/2$ term which appears in
\eqref{eq:ch6peakest1}, and whether there is any physical
explanation for why the coefficient should take on the value of a
half.

\subsection*{Note}

Unknown to the author, this phenomenon has also been simultaneously
investigated in two other works. In \cite{vaga}, Vladislav Vaganov
analyses the behaviour of radiating perfect fluid models in
$d$-dimensional AdS spacetimes; he notes (as we do here) that there
is a significant change in the behaviour of the total mass for $d >
11$ (where it becomes a monotonic rather than oscillatory function
of the central density), and demonstrates this not only for the mass
but also the temperature and entropy.

He also presents a dynamical systems analysis (based on that given
in \cite{uggla}) of the behaviour for a general linear equation of
state, $P(r) = q \rho(r)$, which includes the radiation case. This
analysis complements the numerical results presented here, providing
an analytic derivation of the critical density, which is determined
to be $d_{c} = 10.964\dots$, consistent with our relation
\eqref{eq:ch6peakest1}. The specific analysis for the radiation case
is given in section \ref{sec:dynamsys}, where we give not only the
derivation of the critical density, but also demonstrate how the
dynamical systems technique gives analytical expressions for other
parameters introduced in our numerical investigation into the
self-similar behaviour for $d < d_{c}$.

The second related paper, \cite{chan} by Pierre-Henri Chavanis,
presents an in-depth study of the behaviour of general stars
(``isothermal spheres'') with a linear equation of state in an
asymptotically flat background. His results are again complementary,
finding that there is monotonic behaviour for $d \ge 11$, in
contrast to the oscillatory behaviour observed in lower dimensions.
By asymptotic analysis he also finds the value for the critical
dimension in the radiation case to be very close to eleven, and
although there initially appeared to be a discrepancy between the
two alternative calculations of the critical dimension in
\cite{vaga} and \cite{chan}, the latter was subsequently corrected
to agree with the value of $d_{c} = 10.964\dots$ found in
\cite{vaga}. His paper also includes a comprehensive investigation
into the stability of the different regimes, looking at a number of
alternative stellar configurations and considering the behaviour of
other thermodynamic parameters (entropy, temperature,$\dots$), in
addition to the mass.

\section*{Acknowledgements}

I'd like to thank Veronika Hubeny for useful discussions and ideas,
and Don Page for his comments which helped motivate this work, which
was supported by an EPSRC studentship grant and the University of
Durham Department of Mathematical Sciences.

\end{document}